# Piezoresponse phase as variable in electromechanical characterization

Sabine M. Neumayer[1], Sahar Saremi[2,3], Lane W. Martin[2,3], Liam Collins[1], Alexander Tselev[4], Stephen Jesse[1], Sergei V. Kalinin[1], Nina Balke[1*]

[1]Center for Nanophase Materials Sciences, Oak Ridge National Laboratory, Oak Ridge, TN 37831

[2]Department of Materials Science and Engineering, University of California, Berkeley, Berkeley, CA 94720

[3]Materials Sciences Division, Lawrence Berkeley National Laboratory, Berkeley, CA 94720

[4]Department of Physics and CICECO-Aveiro Institute of Materials, University of Aveiro, 3810-193 Aveiro, Portugal

*balken@ornl.gov

*Piezoresponse force microscopy (PFM) is a powerful characterization technique to readily image and manipulate ferroelectrics domains. PFM gives insight into the strength of local piezoelectric coupling as well as polarization direction through PFM amplitude and phase, respectively. Converting measured arbitrary units to physical material parameters, however, remains a challenge. While much effort has been spent on quantifying the PFM amplitude signal, little attention has been given to the PFM phase and it is often arbitrarily adjusted to fit expectations or processed as recorded. This is problematic when investigating materials with unknown or potentially negative sign of the probed effective electrostrictive coefficient or strong frequency dispersion of electromechanical responses since assumptions about the phase cannot be reliably made. The PFM phase can, however, provide important information on the polarization orientation and the sign of the electrostrictive coefficient. Most notably, the orientation of the PFM hysteresis loop is determined by the PFM phase. Moreover, when presenting PFM data as a combined signal, the resulting response can be artificially lowered or asymmetric if the phase data has not been correctly processed. Here, we demonstrate a path to identify the phase offset required to extract correct meaning from PFM phase data. We explore different sources of phase offsets including the experimental setup, instrumental contributions, and data analysis. We discuss the physical working principles of PFM and develop a strategy to extract physical meaning from the PFM phase. The proposed procedures are verified on two materials with positive and negative piezoelectric coefficients.*

## I. INTRODUCTION

Progress in many areas of science is indelibly linked to advances in techniques to investigate functional behavior on the micro- and nanoscale have become essential in material science and device engineering. Among these characterization techniques in areas such as ferroelectricity, energy storage and conversion, and information technologies much of the advancement is related to the development of piezoresponse force microscopy (PFM) and PFM switching spectroscopy modes. These techniques allow for the study of piezoelectric and ferroelectric activity on the micro- and nanometer scale via detection of mechanical response to electric fields applied to the sample via a conductive probe.[1,2] PFM has become a popular tool within the materials-science community, however, not only is the technique inherently prone to artefacts to be aware of, but correct processing and interpretation of PFM signals is not always straightforward. Some of the PFM challenges, especially for resonance-enhanced PFM, have been resolved or mitigated by developing new methods, *e.g.*, artificial PFM contrast due to resonance peak shifts can be avoided by dual resonance frequency tracking[2] and band excitation.[3] Other challenges involve ruling out electrostatic forces or other non-electromechanical phenomena as the predominant signal origin.[4] In addition, advanced calibration procedures[5] or imaging techniques using laser Doppler vibrometers or interferometric displacement sensing[6] have been developed to gain fully quantitative information on the PFM amplitude. Little attention has been paid, however, to the PFM phase, although this parameter yields valuable information on fundamental physical properties such as the sign of the electrostrictive coefficients and frequency dispersion of electromechanical response.

PFM measures the periodic surface displacement $L$ along the normal of the surface ($z$-direction) as a result of volume changes induced by an AC voltage $V$ applied to the tip. In-plane PFM, where lateral surface displacement is tracked, can also be conducted but will not be discussed in detail here. The surface displacement is measured indirectly through the cantilever deflection $D$ which is proportional to the changes in slope of the cantilever detected via laser beam deflection. With strain $S = L/L_0$ and electric field $E = -V/L_0$, where $V$ is the voltage applied to the SPM tip (and therefore has the opposite direction of $E$) and $L_0$ is the probing thickness, the PFM signal can be used to obtain information about the piezoelectric constant $d$ probed normal to the sample surface[7]:

$$d_{ij} = \left(\frac{\partial S_j}{\partial E_i}\right) = -\left(\frac{\partial L_j}{\partial V_i}\right) = -D_{ac} \quad (1)$$

Note that if the field is applied to the bottom electrode, the negative sign in Eqn. 2 should be ignored.

If the normal of the surface is aligned with the $z$-direction of the sample and the field is purely normal, PFM would probe the piezoelectric tensor component $d_{33}$. Since that is not always the case, PFM is often described as probing the effective piezoelectric constant simply denoted as $d$.

Moreover, in-plane components can contribute to the measured signal. Detailed discussion of PFM signals in the presence of vertical and in-plane field components for materials of arbitrary orientations is discussed elsewhere.[8]

Independent of the way PFM is performed, it gives access to the PFM amplitude $A$ and phase $\phi$ of the dynamic cantilever deflection $D_{ac}$. The PFM amplitude is proportional to the (effective) piezoelectric constant along the normal of the sample surface while the phase has directional information along the same axis. Often, PFM amplitude and phase are combined to yield the so-called mixed response $D_{ac}$ which is defined as $D_{ac} = A \cdot sin(\phi)$ or $D_{ac} = A \cdot cos(\phi)$. While there have been efforts to quantify PFM to extract $d$ from $D_{ac}$,[9,10] the PFM phase is often adjusted to maximize the mixed-PFM response so that $sin(\phi)$ or $cos(\phi)$ equals ±1 and/or according to expected loop orientation. Some notable exception are Chen et al., who calibrate the phase response in electrochemical strain microscopy against PFM phases measured on ferroelectric samples with known polarization and electrostriction coefficients[11] and Bradler et al. who propose a phase calibration method based on numerical calculations.[12]

Many factors determine the measured PFM phase, which often include unknown contributions, occasionally leading to oppositely oriented PFM phase loops in the literature for the same material system.[13] Even measurements using the same sample and microscope can result in opposite loops, as demonstrated in Figure 1. Switching spectroscopy was performed on a standard ferroelectric PZT sample using a soft ($k$ = 4.2 N/m) and a stiff probe ($k$ = 45.6 N/m). While amplitude levels are comparable (Figure 1a), the as-measured phase loops (Figure 1b), and consequently, $D_{ac}$ loops are oppositely oriented dependent on the cantilever force constant. In addition, the uncorrected phase levels for these two loops vary greatly. This is unexpected since the same material property is probed independent on the used probe.

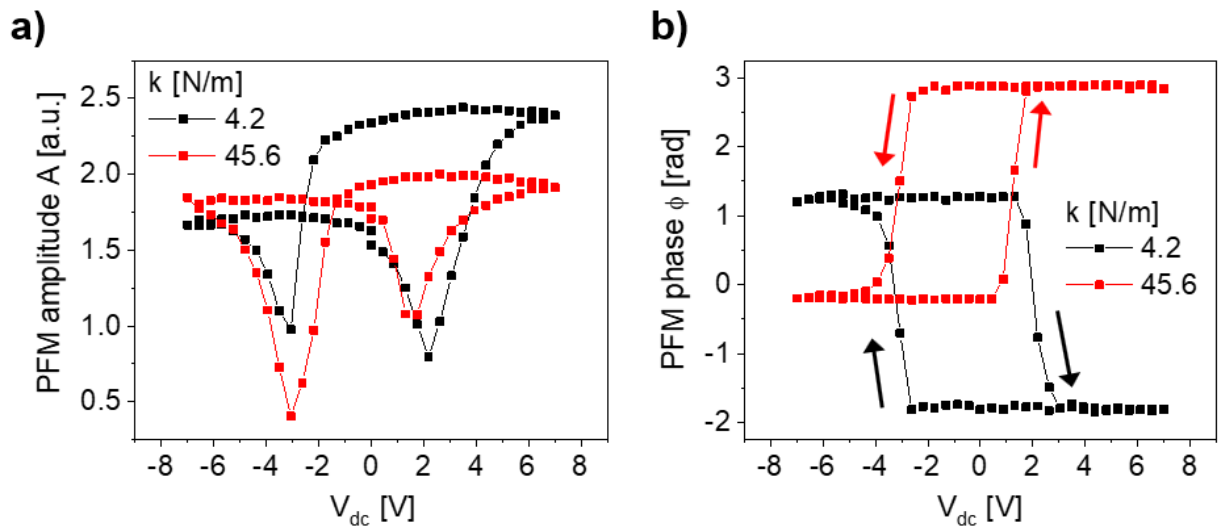

**Figure 1.** PFM (a) amplitude and (b) phase loops measured on the same sample with a soft (4.2 N/m, black) and stiff (45.6 N/m, red) cantilever. The contact resonance frequency for these cantilevers were ~325 kHz and ~625 kHz, respectively. The phase loops show opposite orientations.

If the phase is not correctly processed before calculations of the mixed response, $D_{ac}$ can appear lower than the measured amplitude (*e.g.*, if cos($\phi$) < 1 and > -1) and information provided by the loop orientation is lost or misleading. Therefore, clear guidelines need to be established on how to correctly process the PFM phase if no assumption about loop orientation can be made. This is especially important for materials that exhibit negative electrostriction and therefore also negative piezoelectric coefficients, as for example observed in PVDF[14] and CuInP$_2$S$_6$.[15,16] Similarly, for materials with mobile ions or low-frequency polarization dynamics, the significant dispersion of electromechanical responses can be expected in the range of frequencies amenable to PFM, necessitating quantitative absolute phase measurements.

Here, we establish a systematic approach to correctly identify the PFM loop orientation as determined by the PFM phase. We discuss the role of instrumental phase offsets and their frequency dependence as a main contributor to phase offsets. In addition, the effect of fitting procedures and signal to noise ratio on the extracted PFM phase from simple harmonic oscillator (SHO) fits for on-resonance PFM will be discussed. We will demonstrate how the phase offset can be determined based on the physical meaning of the PFM phase which will be demonstrated on PbZr$_{0.2}$Ti$_{0.8}$O$_3$ (PZT) thin films. The developed procedure will then be applied to CuInP$_2$S$_6$, a ferroelectric material with a negative electrostrictive coefficient[15,16] where the loop orientation is opposite than for traditional ferroelectrics such as PZT.[14,15,17]

## II. RESULTS AND DISCUSSION

First, we discuss how PFM information is processed and how it can be used to establish the sign of the electrostrictive coefficient being probed by PFM measurements. Figures 2a and 2b show the ideal PFM amplitude and phase depicted for a ferroelectric material like PZT. The amplitude has two positive branches which partially overlap, and the PFM phase is hysteretic with a constant phase for each polarization state. A positive electrostrictive coefficient, manifests itself as sample expansion in applied field if the orientation of the polarization vector and electric field vector are parallel. Conversely, the material contracts if the polarization vector and electric field are oriented antiparallel. On the contrary, negative electrostrictive coefficients cause a material to contract if the polarization and electric field are oriented parallel. During PFM, the electric probing AC field is alternating, leading to a periodic sample deformation measured as electromechanical response.

A domain with a downward- pointing polarization direction (and positive electrostriction) therefore oscillates in-phase with the applied electric probing AC field resulting in an electromechanical response in-phase equivalent to $\phi = 0$. On the contrary, a domain with an upwards polarization direction contracts in the positive half cycle of the applied sine probing signal and expands in the negative half, yielding an out-of-phase response equivalent to $\phi = 3.14$ ($\pi$) rad. Therefore, a ferroelectric domain switched downwards with a positive voltage applied to the PFM tip ideally has a PFM phase of $\phi = 0$ when the AC probing voltage is applied to the PFM tip in contact with the downwards oriented domain (red arrow in Figure 2b). A domain switched upwards with a negative voltage ideally has a PFM phase of $\phi = 3.14$ rad (blue arrow in Figure 2b). If PFM amplitude and phase are combined according to $D_{ac} = A \cdot cos(\phi)$, then the corresponding PFM hysteresis loop is oriented counter-clockwise as shown in Figure 1c. According to Equation (1), the loop of the piezoelectric coefficient as a function of voltage is then oriented clockwise (Figure 2d). Here, domains switched upwards show a positive $d$ and domains switched downwards have a negative $d$ owing to opposite directions of the polarization vector.

$$d_{33} = 2\, Q_{33}\, P_3 \varepsilon_{33}\, \varepsilon_0 \qquad (2)$$

According to Equation (2)[18] the sign of the piezoelectric constant is determined by the sign of the electrostrictive coefficient $Q$ and the direction of the polarization vector $P$. From the piezoelectric response loop one can determine the sign of the electrostrictive coefficient if the orientation of the polarization vector is known. For example, applying a negative voltage to the tip switches the polarization vector upwards. Therefore, we can normalize $cos(\phi)$, which contains the directional information of the mixed PFM signal, with the $cos(\phi)$ of the polarization direction representing an upward oriented polarization vector. Consequently, an upward oriented polarization vector is defined as +1 and a downward oriented polarization vector is -1 (Figure 2e). If the piezoelectric constant loop is then divided by this curve, one obtains the sign of electrostrictive coefficient (Figure 2f). Note that $Q$ is a material constant and independent of $V_{dc}$. The reduction in $Q$ during the polarization switching event in Fig. 2f is an artifact due to zero PFM amplitude at the coercive voltages.

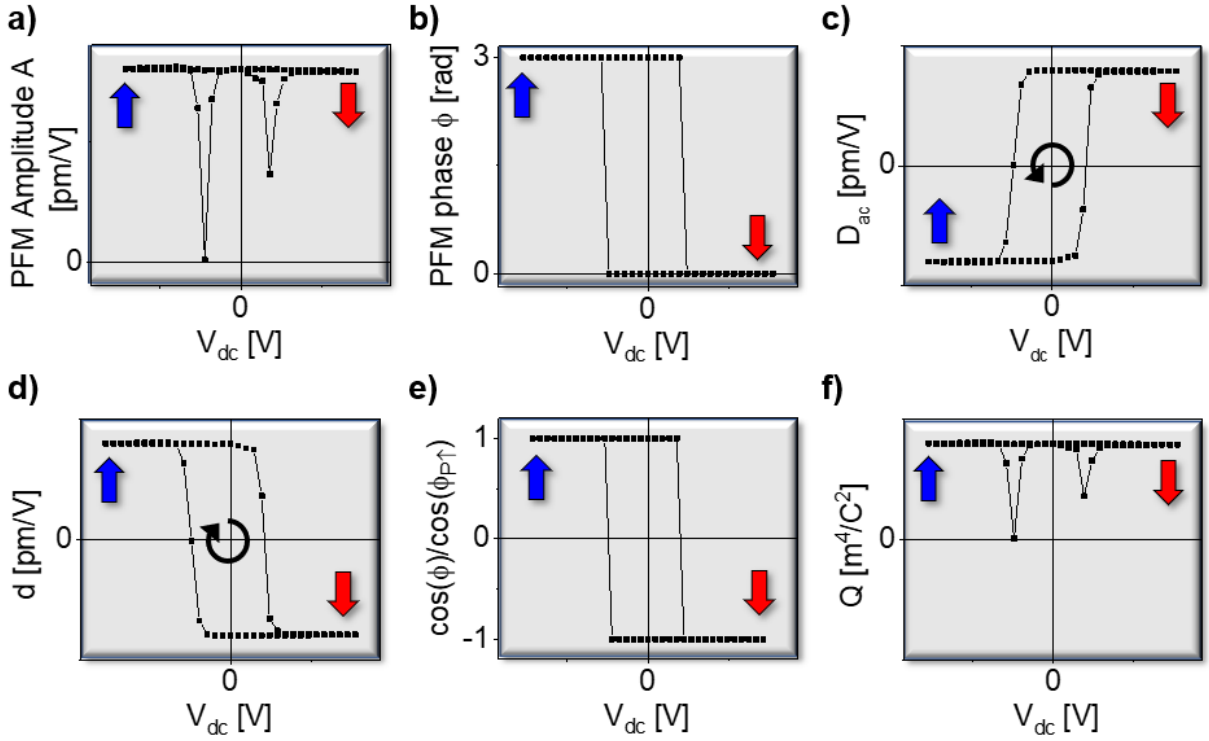

**Figure 2**. Schematic outline of correlation of piezoelectric constant, PFM phase and PFM amplitude hysteresis loop for a material with a positive electrostrictive coefficient. (a) PFM amplitude A, (b) phase $\phi$, (c) mixed PFM signal $D_{ac}$, (d) piezoelectric coefficient $d$, (e) $\cos(\phi)/\cos(\phi_{P\uparrow})$ and (f) electrostrictive coefficient $Q$ as a function of $V_{dc}$ voltage. The blue and red arrow indicates upwards and downwards polarization orientation, respectively.

Practically, the measured PFM phase of oppositely oriented polarization domains is rarely 0 and $\pi$ rad due to instrumental offsets including phase delays in cables, analog signal processing (e.g. amplifiers and filters), cantilever dynamics, and experimental effects such as sample, tip-sample contact, and intrinsic material properties. The instrumental phase offsets can vary significantly and are investigated independent of ferroelectric signal contributions. Towards this goal, we measured the amplitude and phase using internal $V_{ac}$ sources where the output was used as input on the same atomic force microscope (AFM) controller and discovered a strong dependency of amplitude and phase without the presence of a nanoscale probe (Figure 3a and 3b, respectively). This dependence is a common property for all commercial AFMs and also affects all frequency-dependent measurements including non-contact and contact mode cantilever vibrations, as shown in Figure 3c and 3d. We studied the response for several different cantilevers of different stiffnesses by means of cantilever resonances (in non-contact or contact mode) to cover a large frequency range and analyzed the phase to the left of the resonance at the lower end of the frequency window $f_{min}$ (Figure 3c). The result is plotted in Figure 3d and demonstrates that the frequency-dependent instrumental offset is similar for electrostatically driven cantilever vibrations for $SiO_2$ or piezoelectrically driven cantilever vibrations for PZT. Strong linear dependency is observed in both cases and, in the case of PZT, two parallel lines are separated by π and are measured on upwards and downwards oriented domains on PZT.

The origin of the above discussed instrumental phase offset is believed to be instrument electronics, sampling delays, and even the BNC cables. Additional phase offsets can be introduced when cantilever dynamics are involved. Therefore, phase offsets can vary for different microscopes and experimental setups. This results in varying slopes in the phase as function of frequency. The instrumental phase offsets can become so large that measurements with different cantilevers can change the PFM loop orientation. For example, in order to reduce electrostatic signal contributions, it has been suggested to use stiffer tips[10].For resonance-enhanced PFM, however, this increases the frequency of the measurement. Based on the cantilever properties, the higher operating frequency can lead to the complete flip of the PFM phase loop, which explains the opposite loop orientation observed in Figure 1.

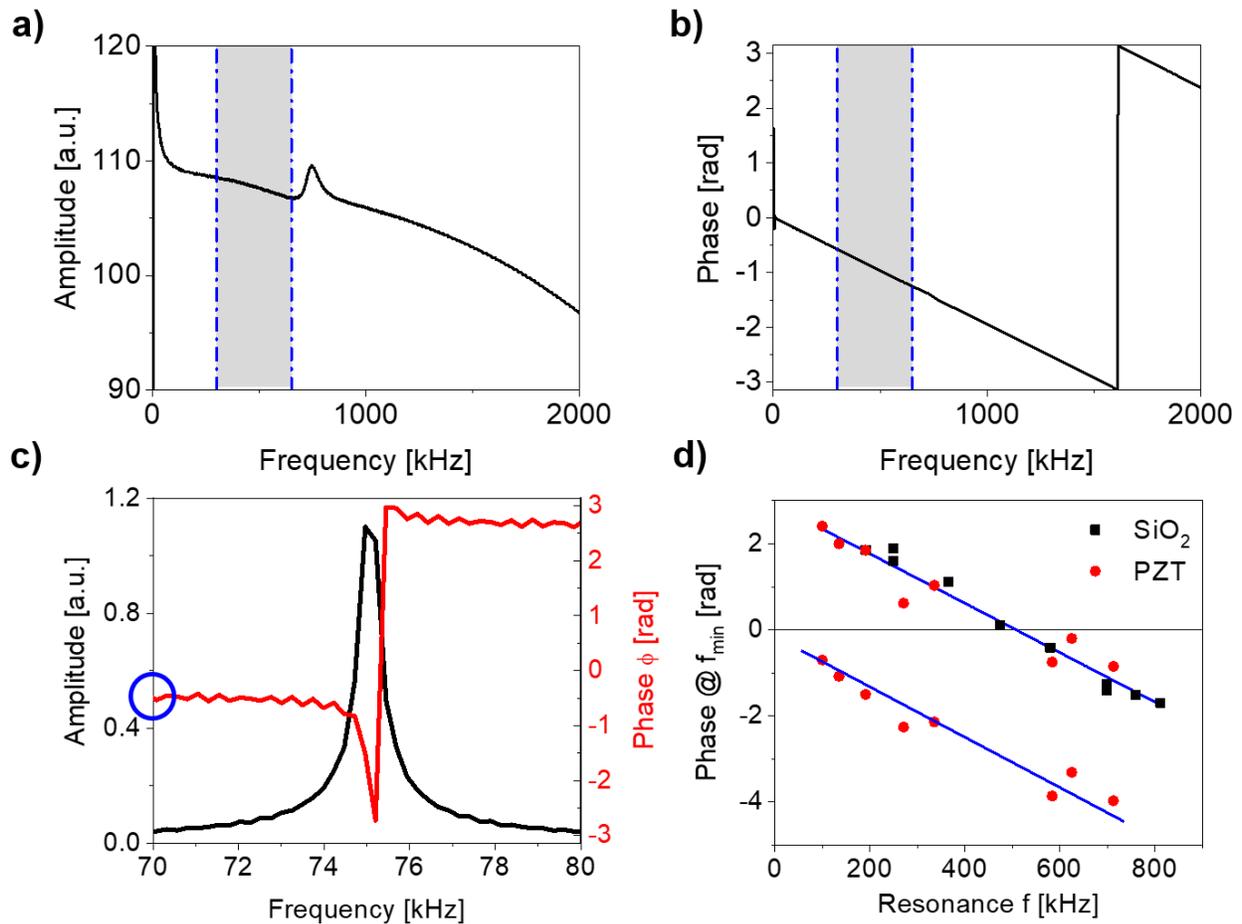

**Figure 3.** Frequency dependence (a,b) purely instrumental in absence of a cantilever and (c,d) as measured in a PFM setup with electromechanical or electrostatic driving. (a) Amplitude and (b) phase as a function of frequency using internal $V_{ac}$ sources which were routed back into the AFM controller as input signal. (c) Oscillating cantilever amplitude and phase across the resonance frequency for an electrostatically driven cantilever above a SiO$_2$ sample (in non-contact mode). (d) Phase at $f_{min}$ (as indicated by blue circle in panel (c)) as a function of frequency for electrostatically driven cantilever vibrations for SiO$_2$ (non-contact and contact mode) and piezoelectrically driven for PZT, collected with many cantilevers of different stiffness. The two bands for PZT, which are separated by π, are measured on upwards and downwards oriented domains.

We would like to emphasize that this frequency dependence of the phase has to be considered for both, on- or off- resonance techniques. However, further considerations for the PFM phase need to be taken into account if measurements are performed at the contact resonance frequency to enhance signal-to-noise ratios.[2,3] In resonance-enhanced techniques, the PFM amplitude and phase are captured across the contact resonance of the tip-sample contact which is determined by mechanical properties of sample and cantilever as well as its geometry. Often, the contact resonance peak is fitted with a simple harmonic oscillator (SHO) equation from which the PFM amplitude $A$, quality factor $Q$, contact resonance frequency $f_c$, and PFM phase $\phi$ are extracted. The quality factor and resonance frequency provide further information on energy dissipation[3] and changes in mechanical properties.[19] The SHO based analysis of phase data centers around three questions: (i) How exactly was the fit performed and the phase extracted? (ii) Is the signal to noise (S/N) ratio high enough for meaningful fitting? and (iii) Does the SHO model adequately describe the measured data?

There are two approaches which are most often used to extract the PFM phase from the SHO fit. First, only the amplitude peak is fitted, and the phase is extracted at the contact resonance frequency (Figure 4a) resulting in a PFM phase value of $\phi$ = 2.67. Second, a 3D-SHO fit of the real and imaginary response as a function of frequency can be performed, where the PFM phase $\phi$ is one of the fitting parameters (Figure 4b) resulting in a PFM phase value of $\phi$ = 1.23. Therefore, different data analysis can result in a different PFM phase value. . Another factor that can influence the SHO fitting parameters, is the S/N ratio. We demonstrate the importance of the S/N ratio for PFM amplitude (Figure 4c) and phase (Figure 4d) by overlaying the data shown in Figure 4b with artificial noise of varying magnitude and subsequent SHO fitting. Both, amplitude and phase clearly become unreliable if the S/N ratio is less than ~3. Therefore, it is important to understand that the PFM phase value depends on how it was extracted from the data and that it is crucial to take into account noise and fit quality. In particular, near coercive voltages, the S/N ratio can significantly decrease due to the low PFM response. Hence, care must be exercised when imposing physical interpretation on fitting parameters $A$, $\phi$, $f_c$ and $Q$ in the vicinity of switching events. In addition to consideration of signal strength and noise levels, it is equally important to understand how well the SHO equation describes the actual resonance. Asymmetric amplitude peak shapes can arise from mechanical non-linearity, electrical cross-talk between drive and response as well as non-piezoelectric signal contribution,[20] which are not captured by the SHO fit (Figure 4e). The same is true for the PFM phase across the resonance which can deviate from π (Figure 4f). This type of behavior can be observed when stiff tips are used to perform PFM experiments on non-ferroelectric samples where electrostatic forces alone drive the cantilever motion. Both, deviations in amplitude and in phase from a SHO model lead to systematic errors in the determination of PFM response.

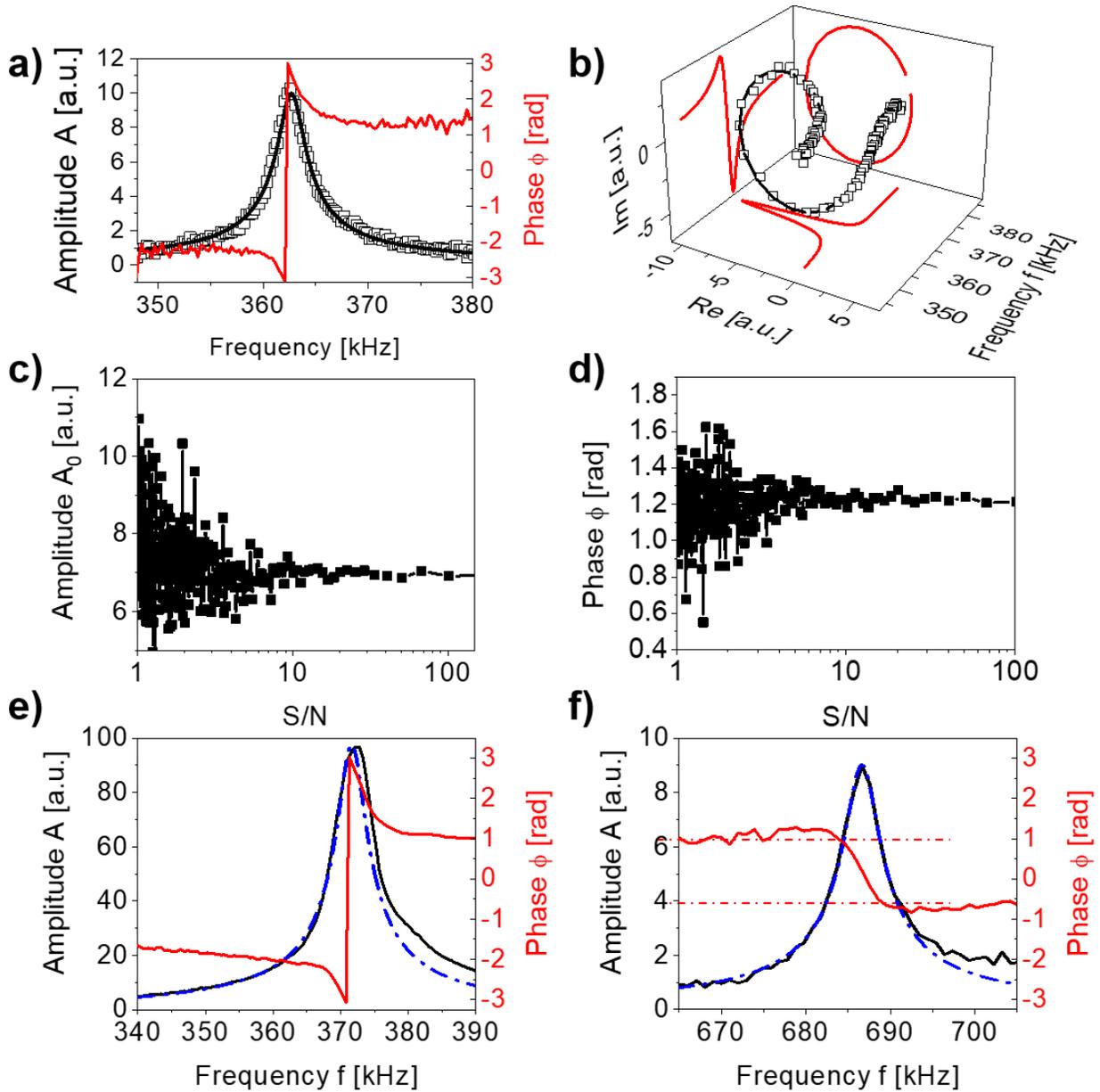

**Figure 4.** Phase and amplitude across the contact resonance in (a) 2D and (b) 3D space with corresponding fitting functions. (c) PFM amplitude and (d) phase extracted from a 3D SHO fit

with artificial noise level. Deviation from a standard SHO in form of a (a) asymmetric amplitude peak and (b) non-π phase shift across the resonance.

If the instrumental phase offset and the exact SHO fitting procedure are unknown, a sample with a known sign of electrostrictive coefficient can be used as reference in combination with systematic rules for PFM phase correction. Once established, the same phase correction offset can be applied to materials with unknown sign of the piezoelectric constant if the measurement set-up and cantilever is not changing. Since PFM contains information about the strength of the piezoelectric response and the direction of polarization, a material with a positive piezoelectric coefficient can show positive or negative values in PFM measurements (Eq. 2). The same is true for materials with a negative piezoelectric coefficient. Therefore, additional information is necessary to uncouple the sign of the piezoelectric coefficient from the polarization direction. If the polarization orientation is unknown or varies locally, poling experiments can help to obtain a reference domain. In the following discussion, all measurements are obtained with cantilevers that have a tip stiffness of 3-4 N/m, and a contact resonance frequency of 330-360 kHz, and all PFM hysteresis loops are measured off-field, which means after application of dc voltage pulses to the tip to initiate domain switching. Here, we performed band excitation PFM imaging after local areas of a 60 nm $PbZr_{0.2}Ti_{0.8}O_3$ /20 nm $SrRuO_3$ /$TbScO_3$ (110) thin film heterostructure have been poled with ± 3 V as indicated in Figure 5a. Here, "positive" or "up" refers to the positive z-direction (pointing away from the substrate) whereas "negative" or "down" refers to the negative z-direction (pointing towards the substrate). A negative voltage applied to the tip in respect to the bottom electrode results in an electric field pointing up which switches the polarization direction upwards while a positive voltage switches it downwards. Based on this principle, we expect the PFM phase in the area poled by applying a negative voltage to have an out-of-phase response with $\phi = \pi$ as discussed above. Conversely, in the positively poled area (polarization orientation down) an in-phase response with $\phi = 0$ (Figure 5b) should be detected. The extracted PFM amplitude and phase plotted as a 2D histogram (Figure 5c) shows PFM phase values of approximately -2.4 rad for the positively poled area and ~0.7 rad for the negatively poled one; neither provide the expected values of 0 and 3.14, respectively. Therefore, the phase has to be offset by $\phi_{offset} = 2.4$ rad and $cos(\phi + \phi_{offset})$ should be used to calculate the mixed-PFM response (Figure 5d).

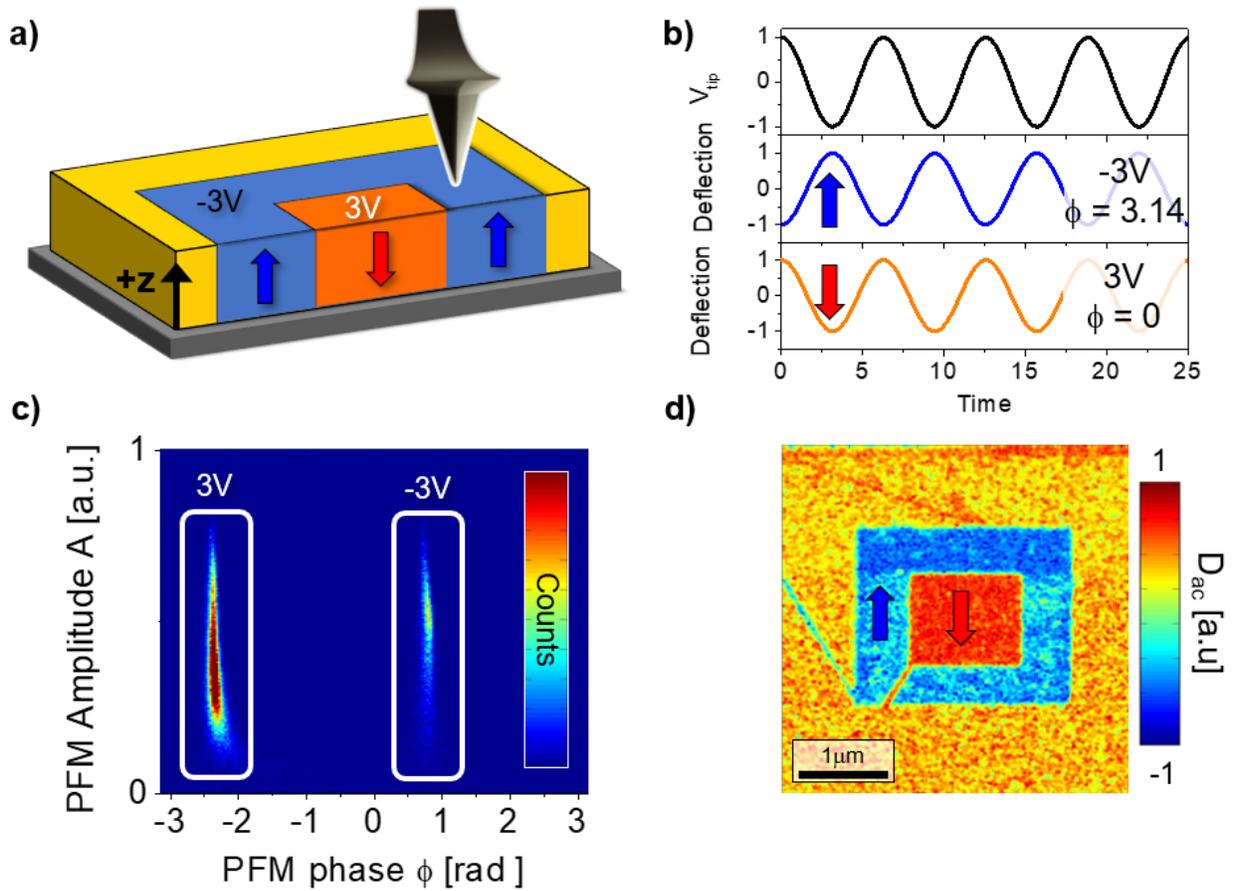

**Figure 5.** (a) Schematic image of the PFM poling experiment (the bias voltage is applied to the probe) and (b) the expected in- or out-of-phase response based on polarization orientation. (c) Measured PFM amplitude and phase plotted as 2D histogram and (d) map of the mixed PFM response in a 5x5 μm² area.

The suggested procedure has been tested on ferroelectric PZT with a positive piezoelectric coefficient as a reference sample and $CuInP_2S_6$ (CIPS) with a negative piezoelectric coefficient (Figure 4). The resulting $d$ loops show the expected loop orientations for the respective sign of the piezoelectric coefficient, *i.e.,* clockwise for positive (Figure 6a) and counter clock-wise for negative $d$ (Figure 6b). It should be emphasized that the outlined considerations about the PFM phase are only valid for the specified frequency at which PFM is measured (330-360 kHz). As previously discussed, if measurements are performed with a stiffer tip at high enough frequencies to cause the phase to change sign, the outlined methodology will need to be adjusted to work in this new frequency range. The same is true if on-field PFM hysteresis loops are measured where additional signal generating mechanisms (such as electrostatic forces) can affect and even reverse the measured phase values. If poling experiments are not feasible, similar approaches can be applied for bias-dependent PFM hysteresis loops. Alternative strategies can include measuring internal phase offsets for mechanically driven cantilever for example through the use of BlueDrive[21] before performing PFM.

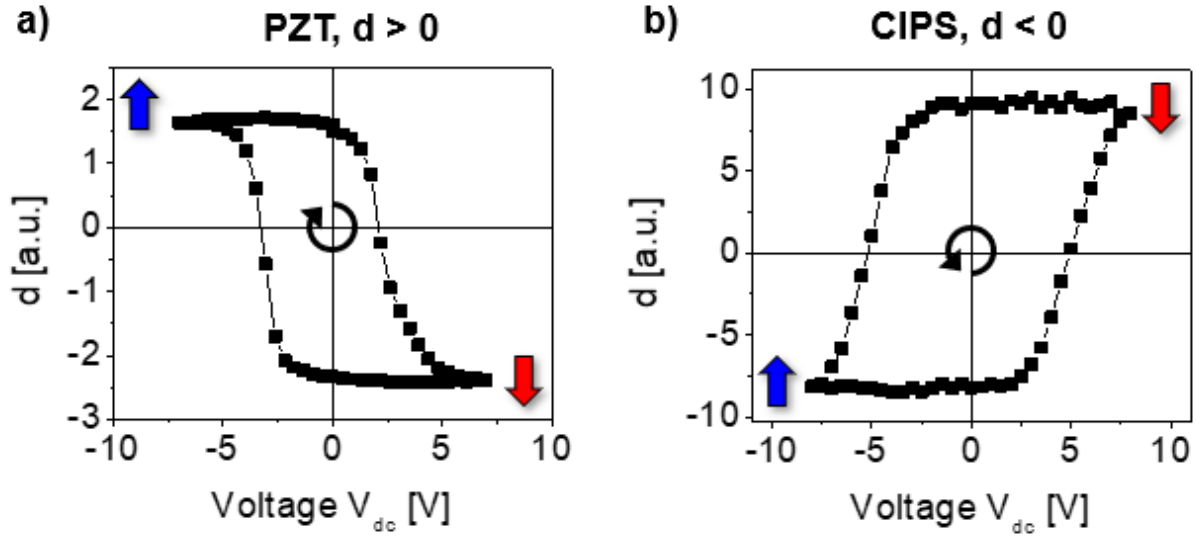

**Figure 6.** Piezoelectric constant loops extracted from PFM response as a function of applied $V_{dc}$ as outlined in this work for a material with a (a) positive and (b) negative piezoelectric constant. (b) is adapted from Neumayer et al. [15]

### III. SUMMARY

In summary, we highlighted the importance of correct processing of the PFM phase signal. PFM phase values deviating from 0 and $\pi$ can artificially reduce the $D_{ac}$ or $d$ response often used to depict PFM hysteresis loops. Moreover, the PFM phase determines the orientation of the PFM loops from which scientific information on the sign of the measured piezoelectric, thus electrostrictive coefficient can be gained. Scientifically accurate processing of the phase signal requires taking into account instrumental phase offsets that occur in virtually all PFM measurements for data processing and interpretation. This instrumental phase offset is dependent on frequency and has to be assessed for each experimental setup separately (since it varies with factors like the stiffness of the cantilever, signal routing, cabling, etc.) In addition, for contact resonance enhanced techniques it is important to take into account the quality of the SHO fit and how the phase data is extracted. Especially in the presence of non-piezoelectric signal contributions, the SHO model might not describe the measured data well enough, *e.g.,* if the peak shape is asymmetric or the phase change across the contact resonance is $\neq \pi$. Moreover, it should be assessed if the S/N ratio is high enough for a proper fit. SHO fitting can be performed in several ways, *e.g.,* fitting only the amplitude peak and extracting the phase at resonance vs. fitting 3D data in imaginary, real, and frequency space where the phase value is a fitting parameter and extracted after resonance. Since the phase varies strongly across resonance it is important to know how/where the phase is extracted and to keep that procedure consistent. In the case presented here, it was necessary to zero the PFM phase level measured on domains with downward oriented polarization vector, so that the phase values are between 0 and 3.14 rad dependent on downwards or upwards polarization orientation, respectively. This convention is in accordance with the working principle of PFM linking periodic in-phase and out-of-phase mechanical deformation in response to an AC electric field. Mixed response $D_{ac}$ should then be calculated as $D_{ac} = A \cdot cos(\phi + \phi_{offset})$ and is coupled with the piezoelectric coefficient $d$ by a factor of -1. If these outlined steps are followed, scientific meaning such as the sign of the piezoelectric constant can be inferred from $D_{ac}$ and $d$ images as well as the orientation of the measured hysteresis loops as a function of applied voltage.

**ACKNOWLEDGEMENTS**

The experiments in this work were performed and supported at the Center for Nanophase Materials Sciences in Oak Ridge National Lab, which is a DOE Office of Science user facility (L.C., S.J., S.V.K., N. B.). The measurements on the PZT sample was supported by the Division of Materials Science and Engineering, Basic Energy Sciences, US Department of Energy. (S.M.N.). S.S. acknowledges support from the National Science Foundation under Grant DMR-1708615 for PZT synthesis. L.W.M. acknowledges support from the Army Research Office under Grant W911NF-14-1-0104 for PZT synthesis. In part (A. T.), this work was developed within the scope of the project CICECO-Aveiro Institute of Materials, FCT Ref. UID/CTM/50011/2019, financed by national funds through the FCT/MCTES.